\documentclass[12pt]{iopart}
\newcommand{\BEQ}{\begin{equation}}     
\newcommand{\BEA}{\begin{eqnarray}}
\newcommand{\EEQ}{\end{equation}}       
\newcommand{\EEA}{\end{eqnarray}}

\renewcommand{\vec}[1]{{\bf{#1}}}       

\begin{document}

\input epsf.sty

\title{Stochastic evolution of four species in cyclic
competition}
\author{C. H. Durney$^1$, S. O. Case$^2$, M. Pleimling$^2$, and R.K.P. Zia$^{2,3}$}
\address{$^1$ Department of Mathematics, Ohio State University, Columbus, Ohio,
43210-1174, USA\\
$^2$Department of Physics, Virginia Tech,
Blacksburg, Virginia, 24061-0435, USA\\
$^3$ Department of Physics and Astronomy, Iowa State University, Ames, Iowa 50011-3160, USA}
\date{\today }

\begin{abstract}
We study the stochastic evolution of four species in cyclic competition in a well mixed environment. In systems
composed of a finite number $N$ of particles these simple interaction rules result in a rich variety of extinction
scenarios, from single species domination to coexistence between non-interacting species. Using exact results
and numerical simulations we discuss the
temporal evolution of the system for different values of $N$, for different values of the reaction rates, 
as well as for different initial conditions.
As expected, the stochastic evolution is found to closely follow the mean-field result for large $N$, with notable
deviations appearing in proximity of extinction events. 
Different ways of characterizing and predicting extinction events are discussed.
\end{abstract}

\maketitle

\section{Introduction}
The study of evolutionary game theory and population dynamics often employs statistical physics and non-linear 
dynamics in order to shed light on multispecies ecological systems \cite{Hof98,Now06,Sza07,Fre09}. 
By searching for generic characteristics 
in simplified models, we can start to understand coexistence and species extinction for complex, real-world systems.  
Many recent studies \cite{Fra96a,Fra96b,Pro99,Tse01,Ker02,Kir04,Rei06,Rei07,Rei08,Cla08,Pel08,Rei08a,Ber09,Ven10,Shi10,
And10,Rul10,Wan10,Mob10,He10,Win10,He11,Rul11,Wan11,Nah11,Jia11,Pla11,Dem11,He12,Van12,Don12,Dob12} have revealed a 
rich behavior in the evolution of three species in cyclic competition.  
This model can be expanded to produce more complex extinction scenarios by simply adding a fourth species
\cite{Fra96a,Fra96b,Kob97,Sat02,Sza04,He05,Sza07b,Sza08,Cas10,Dur11,Rom12,Dob12}. Labeling the different species 
with $A, B, C, D$, we allow $A$ to `prey on' $B$, $B$ to `prey on' $C$, etc.
Interestingly, the four species form non-interacting partner-pairs, much like in the game of Bridge.  
Thus, in a system with $N$ individuals, there are $2(N+1)$ absorbing states, most of which consist of a surviving 
partner-pair: $A$-$C$ or $B$-$D$. This is of course fundamentally different to the three-species case where every
species interacts with every other species.

In \cite{Dur11} we used mean-field theory (MFT) to study the four species situation in absence of spatial and 
temporal fluctuations. MFT trajectories in the four-species case are influenced by a collective variable $Q$, see below, 
which evolves exponentially with an exponent $\lambda$. 
A variety of trajectories in configuration space are encountered, ranging from periodic, saddle-shaped trajectories to
spirals and even trajectories straight as an arrow.
Although MFT is expected to capture some of the stochastic 
behavior, especially for large number $N$ of particles, MFT cannot answer many interesting questions, 
and especially those related to extinction processes. In order to systematically study various extinction scenarios, we 
focus in the following on stochastic evolution by solving the master equation for small number of particles
and by performing Monte Carlo simulations for larger systems. Varying predation rates, initial conditions, and
the number of particles in the system allows us to develop a more comprehensive picture of the stochastic effects
that take place in systems dominated by the formation of alliances of mutually neutral species.

The paper is organized in the following way. We describe the details of the model in Section
\ref{sec:Model}. As the update scheme for the stochastic processes is not unique, we propose two different schemes
and elucidate the relationship between them.
In Section \ref{sec:meanfield}, we summarize the mean-field theory results obtained in \cite{Dur11}
as far as they are relevant for the present study. Section \ref{sec:ExtinctionEvents} is the main part of the paper.
We here describe in some detail different extinction events taking place in our system. 
Solving the master equation, we find exact results for a small system composed of $N = 4$ particles.  
For larger system sizes, we rely on Monte Carlo simulations to study the stochastic evolution.
We study the behavior of $Q$ during the process and discuss how systems may wind up on
the `wrong' $-$ or unexpected $-$ absorbing state.
We also discuss two different maxims that are expected to predict the
probable outcomes of our stochastic processes.
We summarize our findings in the last Section and sketch 
possible avenues for future research.

\section{The model and its absorbing states}
\label{sec:Model}

Our model involves four interacting species with cyclic competition. Denoting particles of each species by $A,B,C$, 
and $D$, the dynamics may consist of picking a random pair (PRP) and, if the individuals are ``{\it cyclically}'' 
different, letting them interact with probabilities $p_a$, $p_b$, etc.:
\begin{eqnarray*}
&&A+B\stackrel{p_a}{\rightarrow }A+A \\
&&B+C\stackrel{p_b}{\rightarrow }B+B \\
&&C+D\stackrel{p_c}{\rightarrow }C+C \\
&&D+A\stackrel{p_d}{\rightarrow }D+D
\end{eqnarray*}
We emphasize that $AC$ and $BD$ pairs are {\it non-interacting}. A mnemonic for these rules could be 
``$A$ consumes $B$ with probability $p_a$; $B$ consumes $C$ ...'' No spatial structure is imposed, so 
the configuration of the system is given by the set of four integers $\left( N_A,N_B,N_C,N_D\right) $
where $N_X$ indicates the number of particles of type $X$ present in the system. 
Note that the total number 
\begin{equation}
N=N_A+N_B+N_C+N_D 
\end{equation}
is an invariant, so that the configuration space is just a set of points within a three-dimensional simplex, 
namely a tetrahedron (of length $N$ on each side).

It is easy to write down the master equation for $P\left( N_A,N_B,N_C,N_D;\tau \right)$, the probability for finding 
the system $\tau$ steps after some given initial distribution, 
$P_0\left(N_A,N_B,N_C,N_D\right) $: 
\begin{eqnarray}
&&P\left(N_A,N_B,N_C,N_D;\tau+1\right) \nonumber \\
&=&\frac{p_a\left( N_A-1\right) \left( N_B+1\right) }{N\left(N-1\right) /2} P\left( N_A-1,N_B+1,N_C,N_D;\tau \right) \nonumber \\ %
&&+\frac{p_b\left( N_B-1\right) \left( N_C+1\right) }{N\left( N-1\right) /2} P\left(N_A,N_B-1,N_C+1,N_D;\tau \right) \nonumber \\ %
&&+\frac{p_c\left( N_C-1\right) \left( N_D+1\right) }{N\left( N-1\right) /2} P\left(N_A,N_B,N_C-1,N_D+1;\tau \right) \nonumber \\ %
&&+\frac{p_d\left( N_D-1\right) \left( N_A+1\right) }{N\left( N-1\right) /2} P\left(N_A+1,N_B,N_C,N_D-1;\tau \right) \nonumber \\ %
&&+\left[1-\frac Z{N\left( N-1\right) /2}\right] P\left( N_A,N_B,N_C,N_D;\tau \right)
\label{PRP-ME}
\end{eqnarray}
where 
\begin{equation}
Z\equiv p_aN_AN_B+p_bN_BN_C+p_cN_CN_D+p_dN_DN_A\,\,.  \label{Z-def}
\end{equation}
{}From here, we can derive a partial differential equation for the generating function. But, to find its solution is 
far from trivial. As summarized in Section \ref{sec:meanfield}, applying a mean-field approach has proven to be fruitful 
in providing insight into the behavior of the system, see \cite{Cas10,Dur11}.

A different approach is to perform computer simulations. However, it is clear that using the PRP scheme, 
many attempts to change the system would
fail and the evolution would be quite slow. To speed up the process, we exploit another scheme, in which we 
always change a pair (ACP) at each step.
Here, ``time'' $t$ is measured in terms of interaction steps. To be precise, if the system at time $t$ consists 
of $N_A\left( t\right) $, $N_B\left( t\right)$, etc., we first construct $Z\left( t\right) $, the combination above. 
We then generate a random number and, if it lies in the range 
\begin{equation}
\left[ 0,\frac{p_aN_A\left( t\right) N_B\left( t\right) }{Z\left( t\right) } \right) \,\,, 
\end{equation}
set $N_A\left( t+1\right) =N_A\left( t\right) +1$ and $N_B\left( t+1\right)=N_B\left( t\right) -1$. Similarly, 
we increase and decrease, respectively, $N_B$ and $N_C$, if it lies in the range 
\begin{equation}
\left[ \frac{p_aN_A\left( t\right) N_B\left( t\right) }{Z\left( t\right) } , \frac{p_aN_A\left( t\right) N_B\left( t\right) }{Z\left( t\right) } + \frac{p_bN_B\left( t\right) N_C\left( t\right) }{Z\left( t\right) } \right) \,\,, 
\end{equation}
etc. In this manner, a configurational change occurs at each increment of $t$,
with the appropriate relative probabilities. The master equation for $
P\left( N_A,N_B,N_C,N_D;t\right) $, can be written for the ACP scheme in the following form:
\begin{eqnarray}
&&P\left( N_A,N_B,N_C,N_D;t+1\right) \nonumber \\
&=&\frac{p_a\left( N_A-1\right) \left( N_B+1\right) }{%
Z+p_a\left( N_A-N_B-1\right) }P\left( N_A-1,N_B+1,N_C,N_D;t\right)  \nonumber \\
&&+\frac{p_b\left( N_B-1\right) \left( N_C+1\right) }{Z+p_b\left( N_B-N_C-1\right) }%
P\left( N_A,N_B-1,N_C+1,N_D;t\right)  \nonumber \\
&&+\frac{p_c\left( N_C-1\right) \left( N_D+1\right) }{Z+p_c\left( N_C-N_D-1\right) }%
P\left( N_A,N_B,N_C-1,N_D+1;t\right)  \nonumber \\
&&+\frac{p_d\left( N_D-1\right) \left( N_A+1\right) }{Z+p_d\left( N_D-N_A-1\right) }%
P\left( N_A+1,N_B,N_C,N_D-1;t\right) ~~~~ .  
\label{ACS-ME}
\end{eqnarray}

However, the presence of variables in the denominators poses serious challenges, even for deriving an equation 
for the generating function. Needless to say, the details of the evolution under these two schemes will be very different. 

Let us remark that each face of the tetrahedron is ``absorbing'' in the sense that transitions into the face 
are irreversible and corresponds to the extinction of one of the four species. Within each face is a special 
limit of the problem with cyclic competition of three species, where one of the three rates is zero. It follows 
that lines between nodes indicate extinction of two species and vertices indicate the extinction of all but one 
species. Further, unlike the cyclic competition of three species, which only has three absorbing states 
regardless of N, our system has $2\left( N+1\right) $ absorbing states, namely, the two fixed lines: $A$-$C$ and $B$-$D$. 
Therefore, it is a priori unclear if the transition probabilities, associated with an initial configuration ending up 
in {\it any particular} absorbing state, are the same for the two schemes. Fortunately, we are able to prove that these extinction 
probabilities are scheme-independent (see the Appendix for details). Thus, we can use the fast, ACP scheme with 
confidence, although we should refrain from comparing every aspect of its full evolution with the time dependence 
from, say, a mean-field approximation of the PRP updates.


\section{Summary and implications of mean-field theory}
\label{sec:meanfield}
While writing the master equation $ \left( \ref{PRP-ME} \right) $ is facile, solving it or obtaining an 
appropriate generating function is very challenging.  However, much insight into this problem can be 
gained by assuming a far simpler approach: the mean-field approximation, which focuses on the mean values 
of the fractions
\begin{equation}
\left\langle X_i\right\rangle_\tau/N\equiv \sum \left( X_i/N\right) P\left(A,B,C,D;\tau\right) \,\,.  \label{X(t)}
\end{equation}
Mean field theory of our four species model was the topic of two recent publications. Therefore, we here only 
summarize the main results that are needed for the following discussion and refer the reader to 
\cite{Cas10,Dur11} for more details.

We find it convenient to simplify the notation by denoting these {\it fractions} by $A(\tau)$, $B(\tau)$, etc. from here 
on and alert the reader when necessity requires us to revert to integer values.  Thus, we have 
\begin{equation}
A\left( \tau\right) +B\left( \tau\right) +C\left( \tau\right) +D\left( \tau\right)
=1\,\,. 
\end{equation}

Following standard routes, we can derive an equation for the changes, 
$\left\langle X_i\right\rangle _{\tau+1}-\left\langle X_i\right\rangle _\tau$, from
equation (\ref{PRP-ME}). These will involve averages of products (e.g., 
$\left\langle AB\right\rangle _\tau$) on the right. The mean-field approximation
consists of neglecting all correlations and replacing the averages of
products by the products of averages, so that the end result is a closed set
of determinstic equations for the averages $A\left( \tau\right) $, $B\left(
\tau\right) $, etc. Finally, by rescaling time with $N$ and considering the 
$N\rightarrow \infty $ limit, $\tau$ can be regarded as continuous and
differences can be replaced by $\partial _\tau$. The result is the mean-field
approximation: 
\begin{eqnarray}
\partial _\tau A &=&\left[ k_aB-k_dD\right] A  \label{ABCD eqns 1} \\
\partial _\tau B &=&\left[ k_bC-k_aA\right] B  \label{ABCD eqns 2} \\
\partial _\tau C &=&\left[ k_cD-k_bB\right] C  \label{ABCD eqns 3} \\
\partial _\tau D &=&\left[ k_dA-k_cC\right] D  \label{ABCD eqns 4}
\end{eqnarray}
where we suppressed the $\left( \tau \right) $ in $A\left( \tau \right) $, etc. In
this form, the probabilities ($k$'s) can be thought of as rates, which are
often normalized to $k_a+k_b+k_c+k_d=1$ in the literature. These rates are related to 
the original probabilities via $k_m=2p_m$ \cite{Cas10}.

Since exponential growth and decay in populations are common, it is natural to write 
equations (\ref{ABCD eqns 1})-(\ref{ABCD eqns 4}) as 
\begin{eqnarray}
\partial _\tau \ln A &=&k_aB-k_dD;\quad \partial _\tau \ln C=k_cD-k_bB  \label{lnAC}
\\
\partial _\tau \ln B &=&k_bC-k_aA;\quad \partial _\tau \ln D=k_dA-k_cC  \label{lnBD}
\end{eqnarray}
which also exposes a coupling between the pairs $AC$ and $BD$, which will often be referred 
to as partner pairs. Constructing appropriate linear combinations, we exploit an important control parameter,
\begin{equation}
 \lambda \equiv k_{a}k_{c}-k_{b}k_{d} \label{lambda}
\end{equation}
which allows us to highlight the role that a single species has on the growth/decay of the opposing 
partner pair, producing
\begin{eqnarray*}
\partial _\tau \left[ k_b\ln A+k_a\ln C\right] &=&\lambda D;\quad \partial
_\tau \left[ k_c\ln A+k_d\ln C\right] =\lambda B \\
\partial _\tau \left[ k_c\ln B+k_b\ln D\right] &=&-\lambda A;\partial
_\tau \left[k_d\ln B+k_a\ln D\right] =-\lambda C .
\end{eqnarray*}
Adding and subtracting appropriately, we see that the quantity
\begin{equation}
 Q \equiv \frac{A^{k_{b}+k_{c}}C^{k_{d}+k_{a}}}{B^{k_{c}+k_{d}}D^{k_{a}+k_{b}}}
\end{equation}
evolves in an extremely simple manner:
\begin{equation}
 Q\left(\tau \right)=Q\left(0\right)e^{\lambda \tau }~.
\end{equation}

We believe $Q$ is a generalization of the quantity $R \equiv A^{k_{b}}B^{k_{c}}C^{k_{a}}$ that has been 
introduced in \cite{Ber09} for the three species case, except that $R$ is {\it invariant} for {\it any} set of rates.  This is not 
the case with $Q$ as it has a time dependence.  Clearly, special properties will be manifested in the class 
with $\lambda = 0$, and we will refer to them as ``quasi-stationary systems.''

For those readers interested in deterministic trajectories, their origin, and in-depth analysis we point 
them to \cite{Cas10}.  Since these results have been well-reported, we choose to only focus on the conclusions 
here.  

\subsection{Periodic systems: $\lambda = 0$}
For such systems, $k_{a}k_{c}=k_{b}k_{d}$ and consequently, the numerator {\it and} denominator of $Q$ 
are both invariant. Thus, $Q$ is an invariant as well.  In the evolution of this system, MFT provides that 
none of the species go extinct and the system evolves periodically. Much like the three species case, there is a 
closed loop in phase space. The closed loop can be described by defining the constants of motion 
\begin{equation}
 f \equiv A^{k_b}C^{k_a}; \quad g \equiv B^{k_d}D^{k_a}, 
\end{equation}
which can be used to define hyperbolic sheets through the tetrahedron, and their intersection is the closed 
loop and resembles (the rim of) a saddle.

To continue our characterization of such an orbit, we consider the extremal points
of the time evolution of $A$, denoted by $\hat{A}_\pm$.  
At such a point, the values of the other three fractions are, in general, not extremal themselves.  As the 
details have previously been reported we remark that the equation for fixing them is
\begin{equation}
 A_0^{-k_b/k_a}\hat{A}_{\pm}+C_0\hat{A}_{\pm}^{-k_b/k_a} = \beta
\end{equation}
where $\beta$ is a constant and depends only on the rates and the initial conditions $B_0$ and $D_0$.  
Once $\hat{A}_\pm$ is known it is a simple 
procedure to obtain the remaining values.  As $\hat{A}_\pm$ varies from $0$ to $1$ there are typically 
two solutions.  When these solutions coincide, $\hat{A}_+=\hat{A}_-$, we are at the {\it fixed line}  
formed by the intersection of the two planes: $k_aA=k_bC$ and $k_aB=k_dD$. Exploiting $A+B+C+D=1$, the 
fixed line is easily represented with the parameter $\gamma \in [0,1]$:
\begin{eqnarray}
\left(A^*,C^*\right) = \frac{\left(k_b,k_a\right)}{k_a+k_b}~\gamma;\quad \left(B^*,D^*\right) = \frac{\left(k_d,k_a\right)}{k_a+k_d}
(1 - \gamma).
\end{eqnarray}
Clearly, the line runs between the absorbing $A-C$ and $B-D$ lines.  

If the system begins in the neighborhood of the fixed line, the trajectory is nearly circular with angular 
frequency $\sqrt{k_bk_d \left(1-\gamma \right)}$ and further away begins to represent the surface of the 
tetrahedron \cite{Dob12}. This provides the saddle shape as described earlier in typical systems, systems 
not in the neighborhood of the fixed line or near an absorbing boundary.

We caution the reader that in this regime ($\lambda = 0$) MFT is a good indicator of the behavior of a 
fully stochastic system only {\it until} an extinction event is near. The aim of Section 
\ref{sec:ExtinctionEvents} is to address this issue.

\subsection{Spirals and arrows: $\lambda \neq 0$}
\label{sec:MFlambdneq0}
With such rates, $Q$ grows/decays exponentially, so that there are {\it no} non-trivial fixed points. 
Since the fractions are bounded by unity, this
implies that either $BD$ or $AC$ vanishes in the large $t$ limit. In other words, for a system with 
finite number of individuals, extinction of one of
the species must occur quite rapidly, see Figure 4a of \cite{Cas10} for an example. For many 
questions of interest, the mean-field approach faces, not surprisingly, serious limitations. 
For example, it cannot predict, given a finite system with stochastic dynamics, the (average) 
time for one of the four species to become extinct, or the time for the system to reach an absorbing 
state. Another interesting question is: At the time when one population goes extinct, what is 
the composition of the other three species. In other words, when the system ``lands'' on an absorbing 
face (of the tetrahedron), where does it land and what is the associated probability. Nevertheless, 
we can pursue its consequences once we are given where we land. For example, suppose $D$ vanishes first 
and we land at $\left(A_i,B_i,C_i\right) $ on the $ABC$ face. Such a three-species system is much 
easier to analyze than those in \cite{Ber09}, since $C$ does not consume $A$. In the mean-field 
approximation, the solution is trivial: From equation (\ref{lnAC}), we see that $A^{k_b}C^{k_a}$ is an 
invariant while $A$ is monotonically increasing ($C$ being the non-consumer). Since the evolution ends 
when $B$ becomes extinct, the endpoint, $\left( A_f,0,C_f\right) $ on the $AC$ line is given by an 
equation for, say, $A_f:$
\begin{equation} \label{eq:ACline}
A_f^{k_b/k_a}\left( 1-A_f\right) =A_i^{k_b/k_a}C_i\,\,. 
\end{equation}
This equation can be solved in general, graphically or numerically. Of the two solutions, the larger must 
be $A_f$ (since $A$ is monotonically
increasing). As will be shown below, these predictions are useful for a finite, stochastic system only 
when $\lambda $ is not too close to zero.

\section{Extinction events}
\label{sec:ExtinctionEvents}

The full stochastic problem of extinction probabilities can be stated as follows: Given a set of rates 
and an initial condition, what is the probability that the system will be found in a stationary final 
state? Furthermore, at what time is a species most likely to go extinct?  And what is the average time 
for extinction?  Needless to say, finding the exact answers to these questions is far from easy, even 
for the three-species model. In our case with four species, the problem is considerably more difficult, 
since there is a ``macroscopically large'' number ($2N+2$) of absorbing states. For typical values of $N$ 
in the hundreds or thousands, even exploring this issue with simulations is non-trivial, since the parameter 
space is six-dimensional for each $N$ (three for the rates and three for the initial fractions).  It is therefore 
unrealistic to study the full parameter space.
In this Section, we will provide insight into the stochastic behavior of a number of interesting cases, 
using both exact solutions and Monte Carlo simulations with ACP rates.
Earlier publications \cite{Cas10} and \cite{Dur11} reported two general maxims concerning extinction of a species: 
``The prey of the prey of the weakest is the {\it least likely} to survive'' and ``The prey of the prey 
of the strongest is {\it quite likely} to survive.''  In the following subsections, we further validate 
these claims, as well as describe their limitations.

\subsection{Exact results for $N=4$ systems}
\label{ExactResults}

Let us first provide exact results for the simplest non-trivial system: $N=4$, starting with one 
individual in each species: $N_A=N_B=N_C=N_D=1$. We will not consider any other initial conditions, all of 
which contain only three species, with relatively trivial extinction probabilities.

In this case, it is more convenient to use the populations, rather than fractions, to denote the system's 
configurations. Thus, the initial state is
given by $\left( 1,1,1,1\right) $. To simplify notation further, we 
set $p_{d}$ to unity and restrict the others rates to the unit cube: $%
p_{a},p_{b},p_{c}\in (0,1]$. This can be done without loss of generality,
since cyclic symmetry allows us to label $D$ as the biggest consumer.

With $N=4$, there are ten extinction probabilities. To compute these
transition probabilities, we simply enumerate all trajectories and their
associated weights. For example, the weight of $\left( 1,0,2,1\right)
\rightarrow \left( 1,0,3,0\right) $ is just $2p_{c}/\left(
2p_{c}+p_{d}\right) =2p_{c}/\left( 1+2p_{c}\right) $. Tabulated below are
the exact transition probabilities from $\left( 1,1,1,1\right) $ to the
final states:
\begin{eqnarray}
\left( 4,0,0,0\right)  &\quad &\frac{2p_{c}\left( 3p_{a}+p_{b}\right)
p_{b}^{2}}{\sigma \left( p_{a}+2p_{b}\right) \left( p_{a}+p_{b}\right)
\left( 2p_{a}+p_{b}\right) }  \nonumber \\
\left( 0,4,0,0\right)  &\quad &\frac{2\left( 3p_{b}+p_{c}\right) p_{c}^{2}}{%
\sigma \left( p_{b}+2p_{c}\right) \left( p_{b}+p_{c}\right) \left(
2p_{b}+p_{c}\right) }  \nonumber \\
\left( 0,0,4,0\right)  &\quad &\frac{2p_{a}\left( 3p_{c}+1\right) }{\sigma
\left( p_{c}+2\right) \left( p_{c}+1\right) \left( 2p_{c}+1\right) } 
\nonumber \\
\left( 0,0,0,4\right)  &\quad &\frac{2p_{b}\left( 3+p_{a}\right) p_{a}^{2}}{%
\sigma \left( 1+2p_{a}\right) \left( 1+p_{a}\right) \left( 2+p_{a}\right) } 
\nonumber \\
\left( 3,0,1,0\right)  &\quad &\frac{4p_{a}^{2}p_{b}p_{c}}{\sigma \left(
p_{a}+2p_{b}\right) \left( p_{a}+p_{b}\right) \left( 2p_{a}+p_{b}\right) } 
\nonumber \\
\left( 2,0,2,0\right)  &\quad &p_{a}p_{c}\frac{p_{a}+2p_{b}+p_{c}+2}{\sigma
\left( p_{c}+2\right) \left( p_{a}+2p_{b}\right) }  \nonumber \\
\left( 1,0,3,0\right)  &\quad &\frac{4p_{a}p_{c}^{2}}{\sigma \left(
p_{c}+2\right) \left( p_{c}+1\right) \left( 2p_{c}+1\right) }  \nonumber \\
\left( 0,3,0,1\right)  &\quad &\frac{4p_{b}^{2}p_{c}}{\sigma \left(
p_{b}+2p_{c}\right) \left( p_{b}+p_{c}\right) \left( 2p_{b}+p_{c}\right) } 
\nonumber \\
\left( 0,2,0,2\right)  &\quad &p_{b}\frac{p_{b}+2p_{c}+1+2p_{a}}{\sigma
\left( 1+2p_{a}\right) \left( p_{b}+2p_{c}\right) }  \nonumber \\
\left( 0,1,0,3\right)  &\quad &\frac{4p_{a}p_{b}}{\sigma \left(
1+2p_{a}\right) \left( 1+p_{a}\right) \left( 2+p_{a}\right) }  \nonumber
\end{eqnarray}%
where
\begin{equation}
\sigma \equiv p_{a}+p_{b}+p_{c}+1~.
\end{equation}%
As an illustration, we consider the most symmetric case: all rates being
unity. Then the transition probabilities to each of the vertices is $1/9$;
to the mid-points of the 2 fixed lines, $1/6$; and to the rest of the points
on the fixed lines, $1/18$.

Also instructive is the case with, say, $p_{b}\ll 1$ (i.e., $B$ being the
\textquotedblleft weakest\textquotedblright\ and $p_{a},p_{c}=O\left(
1\right) $). Then, the final states and the associated probabilities are, to
$O\left( 1\right) $,

\begin{eqnarray}
\left( 0,4,0,0\right)  &\quad &\quad \quad \quad \quad \quad \frac{1}{\sigma 
}  \nonumber \\[0.3cm]
\left( 0,0,4,0\right)  &\quad &\frac{2p_{a}\left( 3p_{c}+1\right) }{\sigma
\left( p_{c}+2\right) \left( p_{c}+1\right) \left( 2p_{c}+1\right) } 
\nonumber \\[0.3cm]
\left( 2,0,2,0\right)  &\quad &\quad \quad \quad p_{c}\frac{p_{a}+p_{c}+2}{%
\sigma \left( p_{c}+2\right) }  \nonumber \\[0.3cm]
\left( 1,0,3,0\right)  &\quad &\frac{4p_{a}p_{c}^{2}}{\sigma \left(
p_{c}+2\right) \left( p_{c}+1\right) \left( 2p_{c}+1\right) }  \nonumber
\end{eqnarray}

A clear conclusion is that the \textquotedblleft prey of its [the weakest,
B] prey\textquotedblright\ has vanishing survival probability. Here, $C$ is $%
B$'s prey so that $D$ is the \textquotedblleft prey of $B$'s
prey\textquotedblright\ -- and goes extinct. Similar to the simple
three-species case (where $A$ survives with probability $p_{b}/\left(
p_{a}+p_{b}+p_{c}\right) $), we arrive at a intuitively reasonable maxim:
The prey of the prey of the weakest is \textit{least likely} to survive. By
contrast, the `law of stay-out' in \cite{Ber09} (\textquotedblleft The
species that is least frequently engaged in interactions has the highest
chance to survive.\textquotedblright ) does not seem to always apply. In
particular, if we let $p_{a}\cong p_{b}\ll 1$ (so that $A$ is the least
interactive species), the probability of $B$ surviving is arguably higher.
To be precise, the probability of \textit{all four} survivors being $B$ is $%
1/\left( 1+p_{c}\right) $ to lowest order, whereas the probability of
\textit{some} survivors being $A$ is $p_{c}/\left( 1+p_{c}\right) $.

\subsection{Stochastic simulations for $\lambda = 0$}
Although it is futile to study the entire parameter space, even for $\lambda = 0$, we can understand the general 
behavior of these unique systems by investigating a limited number of parameters. 

\begin{figure}
\centerline{\epsfxsize=3.25in\ \epsfbox{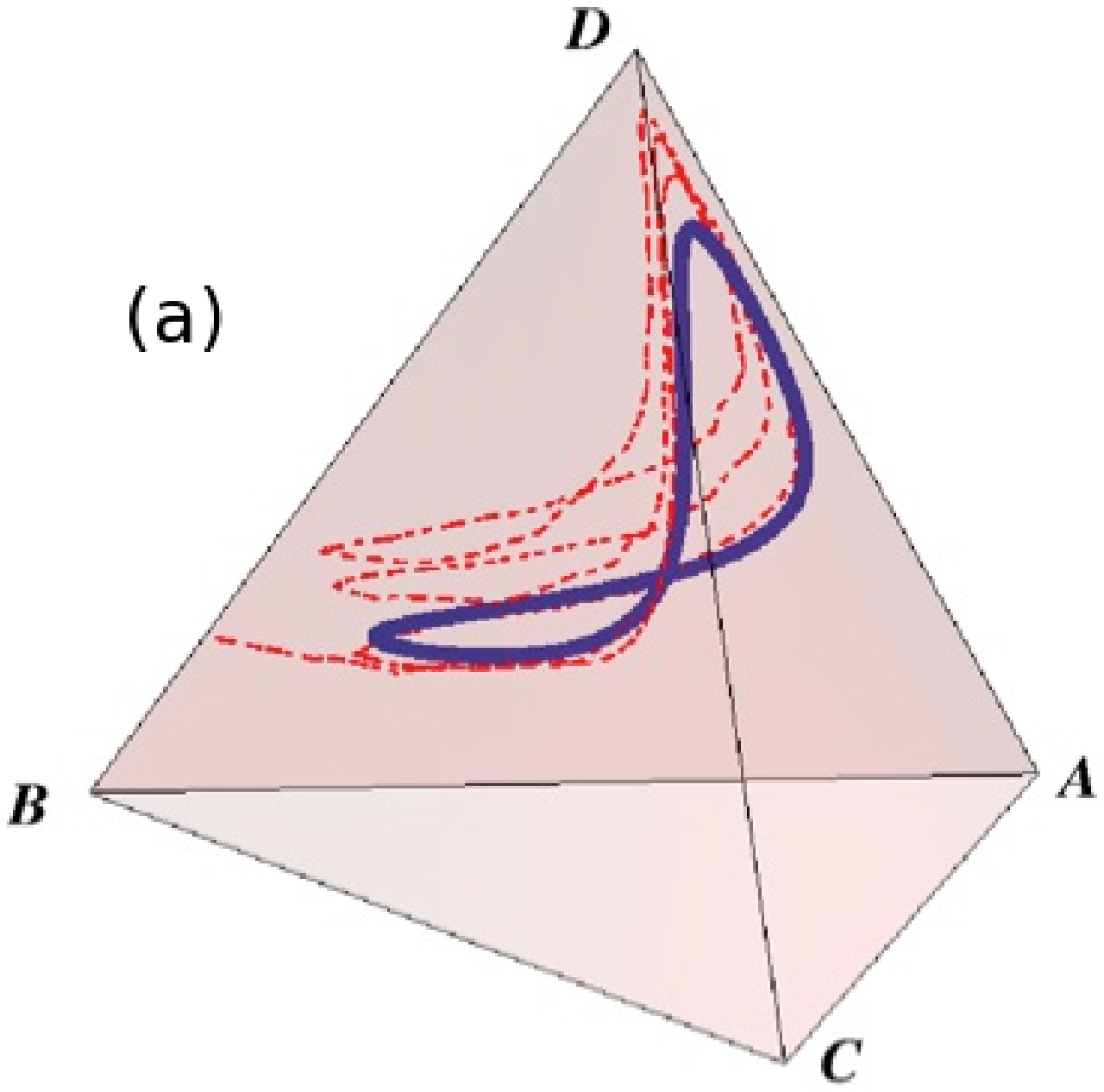} ~~ \epsfxsize=3.25in\ \epsfbox{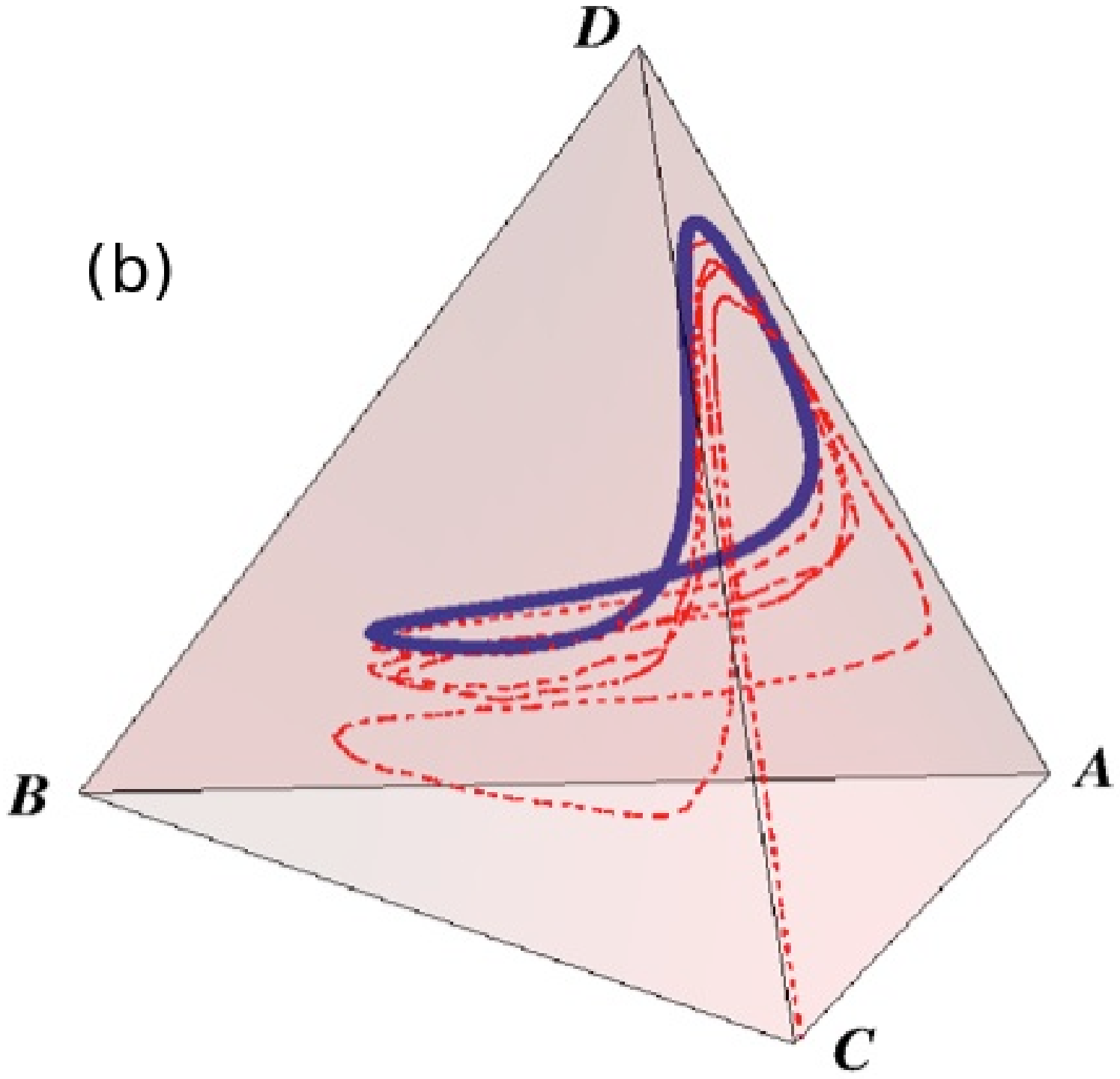}}
\caption{MFT evolution (thick blue line) and Monte Carlo simulation (dashed red line) for initial 
fractions $(A,B,C,D) = (0.02, 0.10, 0.48, 0.40)$ and rates $(k_a,k_b,k_c,k_d) = (0.4, 0.4, 0.1, 0.1)$.  
Here, $\lambda = 0$ and so MFT traces out 
a closed, saddle-shaped orbit.  The simulation initially follows MFT but then, due to stochastic noise, is 
driven to one of the absorbing faces, the system ending either with $B-D$ coexistence (a) or with
$A-C$ coexistence (b).
}
\label{fig1}
\end{figure}

When $\lambda=0$, the mean-field approximation is quasi-stationary, yielding closed loops in the configuration tetrahedron.  
A stochastic trajectory will follow the closed loop at early times, as shown in Figure \ref{fig1}.  
However, due to the intrinsic stochastic noise of the system, the trajectory is driven far from the mean-field trajectory.  
These quasi-periodic trajectories come to an abrupt stop when the trajectories hit one of the absorbing faces of the 
tetrahedron, resulting in the extinction of one species.  In the resulting end game, a second species (the one not 
interacting with the extinct species), dies, yielding a final stationary state with two non-interacting species.  
During different trials of the same initial system, the noise happens randomly and changes the final distribution 
of species dramatically, as exemplified in Figure \ref{fig1} where, despite starting with identical 
parameters, either (a) $BD$ survives or (b) $AC$ survives. 
Consequently, the idea of extinction probabilities does not make much sense with $\lambda = 0$ as extinction is 
based solely on the stochastic noise of the system and clearly the maxims found in earlier publications will not hold.

\subsection{Stochastic simulations for $\lambda \neq 0$}
When $\lambda \neq 0$, mean-field predicts that $Q$ grows/decays exponentially.  Consequently, the stochastic system 
evolves quite rapidly towards an absorbing face and often ends in two-species coexistence. The extinction processes 
happen very early in the stochastic evolution, whereas in the mean-field approximation the system continues to evolve 
in a non-trivial fashion at later times.  For this Section we choose to focus on {\it extreme rates} with an 
{\it asymmetric} initial condition.  The following case is unique, in that the extinction processes only take place in 
very specific windows.  This will allow us to highlight specifically the role that $N \rightarrow \infty$ has on the 
system, as well as to discuss interesting extinction scenarios.

\begin{figure}
\centerline{\epsfxsize=2.55in\ \epsfbox{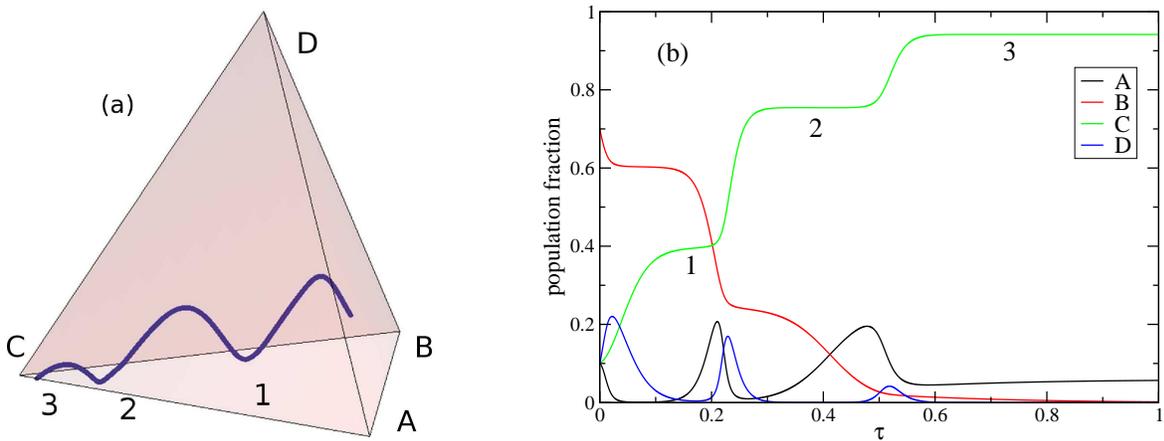} ~~ \epsfxsize=3.25in\ \epsfbox{figure2b.eps}}
\caption{(a) Mean field trajectory for the rates  $\left( k_a, k_b, k_c, k_d \right)=\left( 0.1, 0.0001, 0.1, 0.7999 \right)$
and initial condition $\left( A, B, C, D \right)=\left( 0.1, 0.7, 0.1, 0.1\right)$. The trajectory comes
close to the $ABC$ plane in the regions labeled 1, 2, and 3. (b) Time evolution of the fractions of the different species
for the same case. The data are obtained using a fourth-order Runge-Kutta scheme with time step 
$\Delta \tau = 10^{-5}$.
}
\label{fig2}
\end{figure}

We consider a system with an initial population $\left( A, B, C, D \right)=\left( 0.1, 0.7, 0.1, 0.1\right)$ and  
rates $\left( k_a, k_b, k_c, k_d \right)=\left( 0.1, 0.0001, 0.1, 0.7999 \right)$, resulting in a positive 
$\lambda$. As expected, mean-field theory predicts the evolution of this system to spiral towards the $A-C$ line, see
Figure \ref{fig2}a.  In two 
distinct cases prior to the $t \rightarrow \infty$ limit, the trajectory comes {\it very} close to the $ABC$ absorbing 
face (i.e. $D$'s extinction). The time evolution of the fraction of the different species in mean-field approximation is
shown in Figure \ref{fig2}b. In the time intervals when $D$ is 
close to extinction, the fraction of species $C$ is at a plateau.  
Once the $D$ individuals recover, the number of $C$ individuals also increases, which limits the number of $D$s in the 
system and subsequently leads to another decrease of the number of $D$s. 

It remains unclear to us how to use in the four-species case a methodology 
similar to that proposed in \cite{Rei07} for the three-species case that allows to determine {\it how close} the 
system comes to the absorbing face. As outlined in 
\cite{Rei07} we believe that for four species, the distance to the absorbing face is an extinction probability and 
scales with the number of competing individuals, $N$, but it is an open problem how to approach that
problem analytically.

\begin{figure}
\centerline{\epsfxsize=6.25in\ \epsfbox{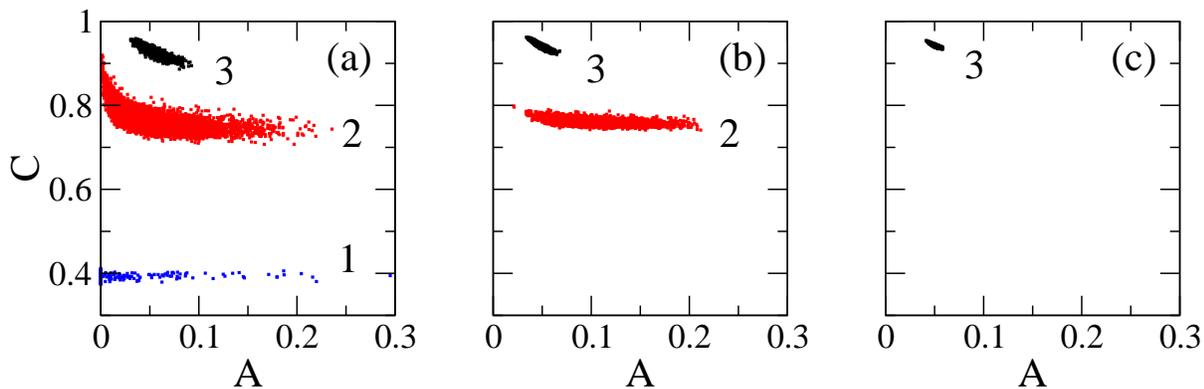}}
\caption{Fractions of $A$ and $C$ particles the moment the $D$ species becomes extinct, as obtained in Monte
Carlo simulations of finite systems, with rates 
$\left( k_a, k_b, k_c, k_d \right)=\left( 0.1, 0.0001, 0.1, 0.7999 \right)$
and initial condition $\left( A, B, C, D \right)=\left( 0.1, 0.7, 0.1, 0.1\right)$. The total number
of particles are (a) $N = 10^4$, (b) $N =10^5$, and (c) $N = 10^6$. The clusters labeled 1, 2, and 3 
correspond to the regions in Figure \ref{fig2} that have the same labels. For each case 10000 independent
runs were done.
}
\label{fig3}
\end{figure}

\begin{figure}
\centerline{\epsfxsize=5.25in\ \epsfbox{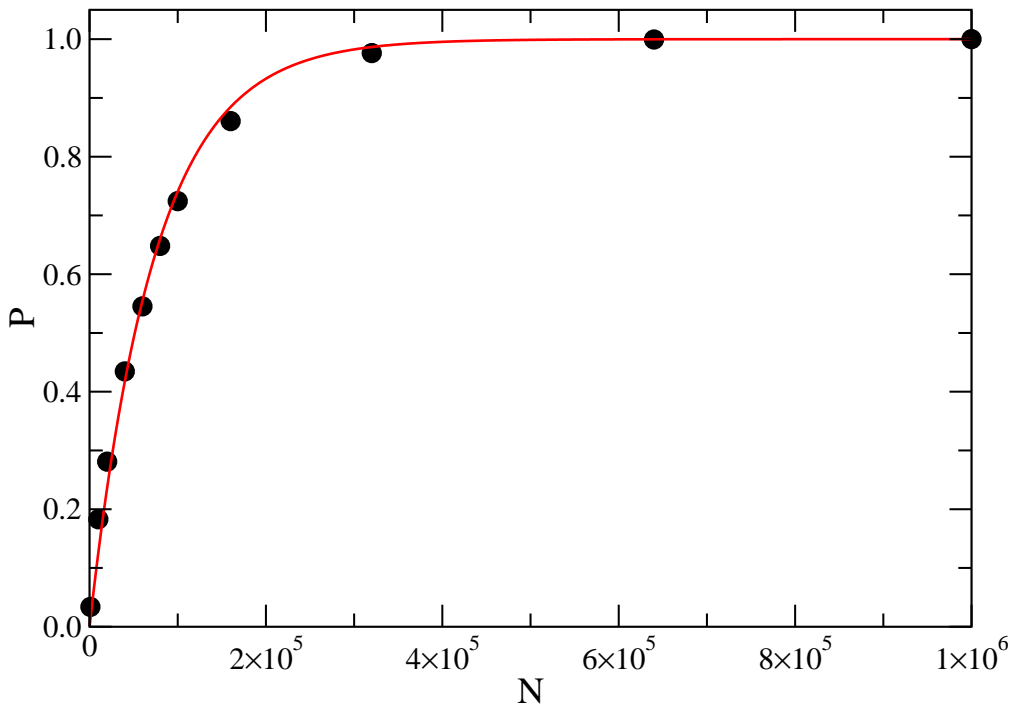}}
\caption{Probability as a function of $N$ that a system ends in the final cluster 3. The rates are 
$\left( k_a, k_b, k_c, k_d \right)=\left( 0.1, 0.0001, 0.1, 0.7999 \right)$
and the initial condition is $\left( A, B, C, D \right)=\left( 0.1, 0.7, 0.1, 0.1\right)$. 
The line is a fit to the function $P = 1- e^{-\alpha N}$, where $\alpha \approx 13.5\times 10^{-6}$.
}
\label{fig4}
\end{figure}

Instead of an analytical approach, we therefore turn to Monte Carlo simulations in order to determine the 
behavior of the stochastic system. In these simulations extinction events
exclusively take place in the time segments where the trajectory is near the $ABC$ face, so that
the fraction of $D$ is very low and the fraction of $C$ is constant.
In Figure \ref{fig3} we discuss three different cases characterized by
different numbers of particles in the system: $N=10^4$, $10^5 $, $10^6$. For each system, we run 10000 independent
simulations using the ACP scheme and plot the $A-C$ composition at the moment when $D$ becomes the first 
extinct species (the number of $B$s follows directly from $A+B+C=1$).  In the smallest system $\left( N = 10^4 \right)$, 
we see three distinct clusters, denoted by 1, 2, and 3. The three clusters correspond to the 
three times the mean-field trajectory comes near the absorbing face, see Figure \ref{fig2}.  Increasing the number 
of particles by 10, cluster 1 completely vanishes and clusters 2 and 3 become better defined.  
Finally, in the $N=10^6$ case, only cluster 3 remains. It is this last cluster that evolves into the
unique long-time mean-field solution in the limit $N \longrightarrow \infty$. The extinction events
that lead to the two other clusters, however, are due exclusively to stochastic effects taking place in
finite systems. As shown in Figure \ref{fig4} the probability $P$ for a given run to end up in cluster 3
varies as a function of system size $N$ as $P = 1- e^{-\alpha N}$, with the numerically determined value 
$\alpha \approx 13.5 \times 10^{-6}$.
This result validates the claim that the distance to the absorbing face scales with $N$.

Finally, let us briefly discuss how our system evolves once the $D$ particles have died out.  
At this point, we are left with a three-species system where the number of $A$ particles, which 
do not have anyone preying on them anymore, increases steadily.  Concurrently, the population of 
$B$ particles also decreases steadily, as they are eaten at a rather high rate by the $A$s, but 
only reproduce with the very small rate $k_b$. In fact, this endgame is very well described by 
mean-field theory, and the time evolution of the $A$ and $C$ populations closely follows the 
mean-field scenario described in Section \ref{sec:MFlambdneq0}: $A^{k_b}C^{k_a}$ is approximately an 
invariant and the positions of the endpoints on the fixed line nicely agree with Equation (\ref{eq:ACline}).

\subsection{The evolution of Q}

As already mentioned previously, the quantity 
\begin{equation}
 Q = \frac{A^{k_{b}+k_{c}}C^{k_{d}+k_{a}}}{B^{k_{c}+k_{d}}D^{k_{a}+k_{b}}}
\end{equation}
fully characterizes the fate of our system in mean-field approximation. Indeed, $Q$ then
grows or decays exponentially according to $Q\left(\tau\right)=Q\left(0\right)e^{\lambda \tau}$
where $\lambda$ is the combination of predation rates given in Equation (\ref{lambda}).  

Before we can directly compare the stochastic evolution of $Q$ to these mean-field results, we first 
need to scale the simulation time $t$ to the mean-field time $\tau$. The rescaling is achieved by
incrementing at a particular time $t$ (the ACP time) the PRP time $\tau$ by
\begin{equation} \label{eq:tauScale}
d\tau(t) = 1/Z(t)
\end{equation}
such that
\begin{equation} \label{eq:tauScale2}
\tau(t+1) = \tau(t) + d\tau(t) .
\end{equation}
Here $Z(t)$ is given by the normalization (\ref{Z-def}).

\begin{figure}
\centerline{\epsfxsize=5.75in\ \epsfbox{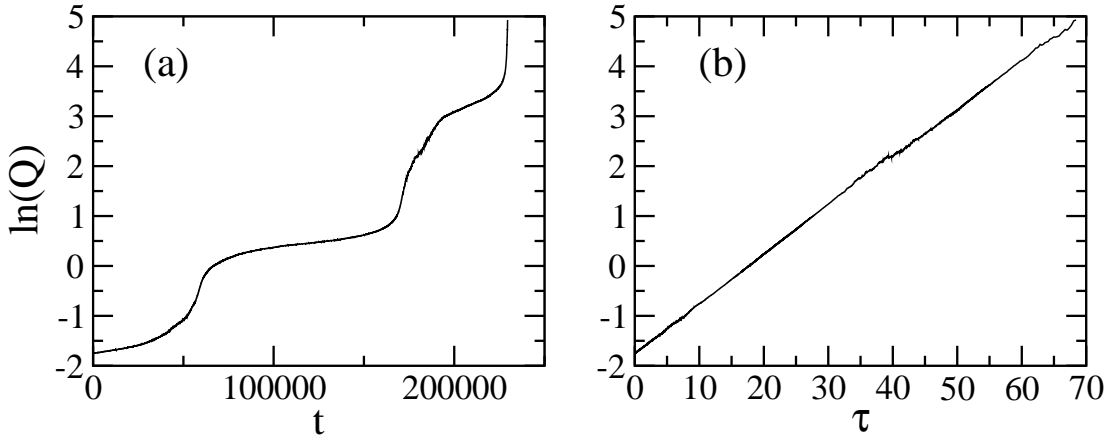}}
\caption{Evolution of $\ln(Q)$ versus (a) ACP time $t$ and (b) PRP time $\tau$ for a system composed of $N = 10^5$ particles,
with rates $\left( k_a, k_b, k_c, k_d \right)=\left( 0.1, 0.0001, 0.1, 0.7999 \right)$
and initial condition $\left( A, B, C, D \right)=\left( 0.1, 0.7, 0.1, 0.1\right)$. 
When plotted as a function of $\tau$, $\ln(Q)$ linearly increases until reaching positive infinity when $D$ dies out (not shown).  
The final distribution of species for this run is 6\% $A$ and 94\% $C$.
}
\label{fig5}
\end{figure}

The difference between the evolution of $Q(t)$ and $Q(\tau)$ is striking, as shown in Figure \ref{fig5}.  
Only after rescaling time, do we find that $\ln(Q)$ evolves roughly linearly with a slope close to $\lambda$.  
Of course, $\ln(Q)$ evolves more linearly for larger system sizes, as is expected from our earlier discussion 
on finite size effects.  Despite this, $Q$ itself is not a reliable indicator of full extinction scenarios.  
As soon as one species dies, $Q$ becomes trivial but competition between the other three species continues.  
Other limitations of $Q$ are exemplified in the following discussion.

\subsubsection{Systems with $\lambda = 0$.}

\begin{figure}
\centerline{\epsfxsize=5.25in\ \epsfbox{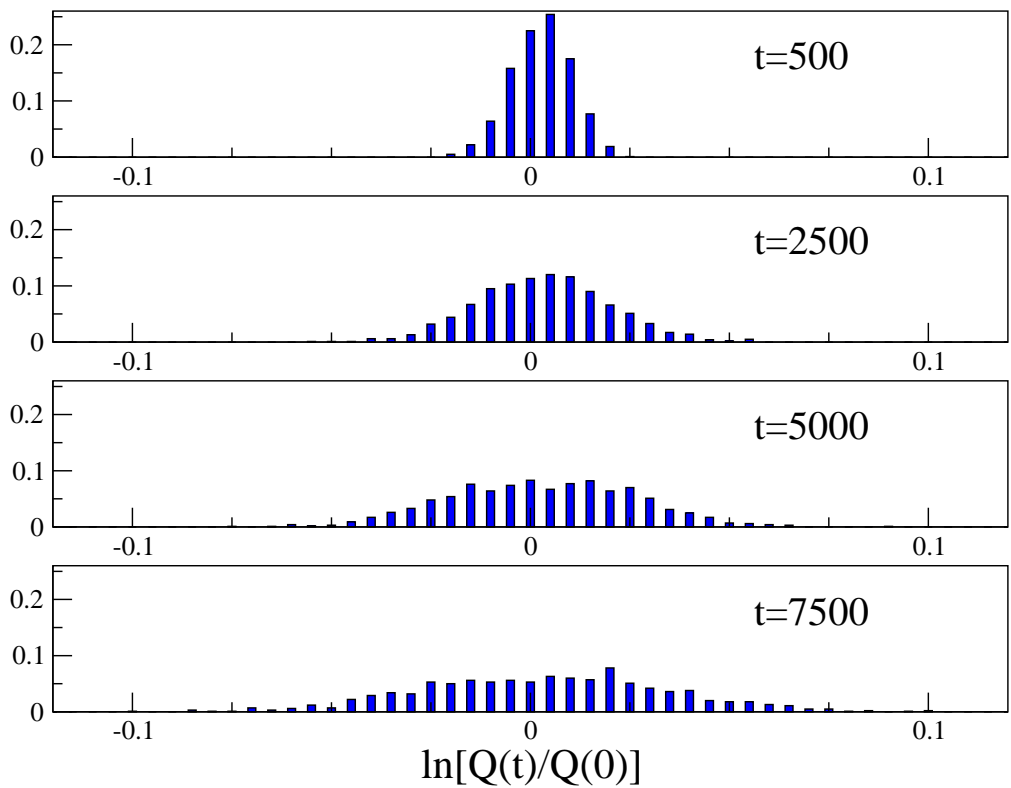}}
\caption{Probability distribution of $\ln(Q(t)/Q(0))$ as a function of ACP time $t$. 
1000 independent runs were done, where A(0) = B(0) = C(0) = D(0) = 2500 and the rates are (0.4, 0.1, 0.1, 0.4), 
making $\lambda$ zero.
}
\label{fig6}
\end{figure}

In mean-field theory, when $\lambda$ is zero, both the numerator and denominator of $Q$ (and, therefore, $Q$ itself) 
are invariant.  However, in the stochastic process, when $\lambda$ is zero, $Q$ wanders from its initial value 
due to stochastic noise. Figure \ref{fig6} illustrates the spreading histogram of $\ln(Q(t)/Q(0))$ 
as a function of the number of interactions.  
This distribution highlights the random fluctuations in $\ln(Q)$ that
reflect the simulation trajectories wandering away from the closed saddle-shaped orbit predicted by mean-field theory.
The spread in Figure \ref{fig6} appears symmetrical, as expected for a purely random distribution.  
However, for a finite system this changes dramatically once extinction events take place.
In fact, over $90\%$ of the trials, that compose the distribution shown in Figure \ref{fig6}, ended with $BD$ 
coexistence despite the seemingly unbiased spread of $\ln(Q)$!

\subsubsection{Systems with $\lambda \neq 0$.}

When the pair $AC$ has a larger product of consumption rates than $BD$, $\lambda$ is positive and, 
consequently, $Q$ grows exponentially, tending towards positive infinity as $B$ or $D$ approaches extinction.  
The opposite occurs when $\lambda$ is negative: $Q$ exponentially decreases until reaching zero when $A$ or 
$C$ dies out. However, there are often exceptions to these generalities.
For example, if $\lambda$ is positive, $Q$ will exponentially increase; but, 
if $A$ or $C$ is the first species to die out, then $Q$ will immediately go to zero and the simulation will 
abandon MFT. The same phenomenon occurs when $\lambda$ is negative and $B$ or $D$ dies first.  
Of course, the final value of $Q$ depends on the population size $N$. Stochastic systems with small $N$ are 
subject to finite size effects and are more likely to stray from MFT.

A good example is provided by our ``extreme rates", with initial population fractions $(0.1, 0.7, 0.1, 0.1)$.
In that case, $\ln(Q)$ initially grows toward positive infinity, as expected for $\lambda > 0$.
However, for this initial condition it sometimes happens for small numbers of particles that 
the first extinction event is given by
the dying of $A$, yielding $B$ as the sole survivor. 
If that is the case, then $\ln(Q)$ jumps to negative infinity, its final value, 
despite its initial increase. In this sense, the stochastic evolution ended on the `wrong'--or unexpected--absorbing face.


\subsection{The maxims revisited}

As the parameter space for systems with three or more cyclically competing species is so huge, attempts have been made
to summarize the possible outcomes by ``laws'' or ``maxims.'' For example, for the three-species case a ``law of the weakest''
was proposed \cite{Ber09} that states that for the case of asymmetric interactions the ``weakest'' species (i.e. the predator with
the smallest predation rate) survives with a probability that approaches one when the number of individuals gets large.
This counter-intuitive observation is in fact due to the unique constellation of three species that all interact with each 
other. Based on our earlier work \cite{Cas10,Dur11}, we proposed the following general maxim: ``The prey of the prey of the weakest is the
{\it least likely} to survive.'' Applying this maxim to the three-species case immediately yields the ``law of the weakest.'' However,
there are additional cases where the outcome of the simulations completely contradict the mean-field prediction. In order
to deal with these cases, the following second maxim was proposed: ``The prey of the prey of the strongest is most likely
to survive.''

In order to get a better grasp at the validity of these maxims, we undertook a systematic study where we fixed the number of
particles ($N=400$) and the initial fractions $(0.25, 0.25, 0.25, 0.25)$, but varied the predation rates. We thereby always set
the largest predation rate (that of species $D$) to $k_d = 1$. In addition to the ``extreme rates'' discussed earlier,
we considered the following values of $\lambda = k_{a}k_{c}-k_{b}k_{d}$: $-0.64$, $-0.16$, $-0.04$, $-0.01$, 
0, 0.01, 0.04, 0.16, and 0.64. For 
every value of $\lambda$, multiple cases were studied, where we let the system run until a stationary state was reached. 
At least $5000$ independent trials were done for every case, and
the survival probabilities were computed. 

Based on these data, we remark that the first maxim faithfully predicts the outcome for our small system
when $\left| \lambda \right|$ is large, i.e. $\left| \lambda \right| = 0.16$ or 0.64. The fate of the system in those cases
is determined by $\lambda$, and only in extremely rare occasions do we observe any deviations from the mean-field
prediction. One such example is the case $(k_a,k_b,k_c,k_d) = (0.1, 0.25, 0.9, 1)$ where in 3\% of the trials
$B$ and $D$ die out. From our previous discussion of finite-size effects, we expect that for larger system sizes
less and less trials will end up in the ``wrong'' steady state. The second maxim also does very well for large
negative $\lambda$ values, where it is by and large equivalent to the first maxim. For large positive $\lambda$,
however, the second maxim miserably fails. As the largest predation rate is $k_d =1$, large positive values of $\lambda$
mean that both $k_a$ and $k_c$ are rather large, whereas $k_b$ is very small. One example is given by
$(k_a,k_b,k_c,k_d) = (0.7, 0.19, 0.5, 1)$, yielding $\lambda = 0.16$. As a result, $A$ preys very efficiently
on $B$ whereas at the same time only few additional $B$s result from the preying of $B$ on $C$. Consequently, $B$
has a very small probability to survive.

For $\lambda$ close to zero, stochastic effects get more and more important, and there is an increasing probability 
that the system does not wind up in the stationary state predicted by mean-field theory. Consequently, the maxims
get less reliable the closer to zero $\left| \lambda \right|$ is. Of course, for an increasing number
of individuals we expect the first maxim to remain valid even close to $\left| \lambda \right|=0$.

\section{Summary and Outlook} 
We investigate the time dependent evolution and extinction probabilities of a simple model 
of population dynamics: $N$ individuals of four different `species', which compete cyclically. Though 
seemingly a trivial extension from a similar three-species game (often called rock-paper-scissors game), this system displays 
a much richer behavior. Since the four species form `partner pairs', the stationary states consist of one 
of the pairs, with $N+1$ possible compositions in each case. As a result, there are $2(N+1)$ absorbing states 
with generally non-trivial distributions among them.
In previous publications, we focused on mean-field theory to study the deterministic evolution of this 
nonlinear system. By manipulating the coupled differential equations, we found a single parameter $\lambda$ 
which controls the system's general behavior and the exponential evolution of $Q$. Of course, in MFT, 
no species ever goes extinct and therefore, we cannot study specific extinction scenarios. Thus, in this 
paper, we focused on stochastic models to explore extinction probabilities and timing. For small $N$, we  
solved the given Master Equation and gave the exact transition probabilities 
for one of these systems, where  $N = 4$ and there is initially one individual of each species.

For large $N$, we have to rely on Monte Carlo simulations. However, the full parameter space is seven-dimensional: 
three parameters for the predation rates (with $k_a + k_b + k_c + k_d = 1$), one for the population size 
$N$, and three for the initial population fractions (with $A + B + C + D = 1$).  With so many variables, 
it is futile to explore the full parameter space, and we focused on a limited number of interesting cases.  
In many cases the systems evolve intuitively. In the limit ${N\to\infty}$, the stochastic 
evolution closely follows the mean-field prediction, as demonstrated in systems with ``extreme rates'' 
and $N = 10^6$. Exploring in more detail the parameter space, we identified the following two maxims: 
``The prey of the prey of the weakest is the {\it least likely} to survive" and ``The prey of the prey of 
the strongest is {\it quite likely} to survive." These maxims, however, do have limitations.
For example, systems with $\lambda = 0$ follow the mean-field orbits at earlier times, but eventually 
wander away from the MFT trajectories and land on a random absorbing face due to the stochastic noise.  
These purely stochastic cases can not be described by maxims.

Simply adding a fourth species to the popular three-species `rock-paper-scissors' game reveals rich, complex 
nonlinear behavior and a tendency towards coexistence. The four-species case is characterized by the formation
of two alliances composed of mutually neutral partners. Whereas we expect similar results for an even number
of species that interact cyclically, thereby yielding two competing alliances, a more complicated situation
prevails for an odd number of species. Among those cases, the three-species
case is very special, as it is the only one where every species interacts with every other. It is therefore an
interesting question what new phenomena emerge for five species, for example. Further interesting
results can be obtained by putting these systems on one- or two-dimensional lattices. Beyond the straightforward
situation of uniform interaction rates, one can imagine cases that are more representative of actual ecological systems
where a prey actively moves away from its predators or where predators strategically chase their prey.
Work along these directions is in preparation.

\ack
We are grateful to E. Frey, K. Mallick, S. Redner, B. Schmittmann, E. Sharpe, and Z. Toroczkai for illuminating discussions. This work is supported in part by the US National Science Foundation through Grants DMR-0705152, DMR-0904999 and DMR-1005417.

\appendix

\section{Equivalence of PRP and ACP schemes}

Let $\vec{x}\equiv \left( A,B,C,D\right) $, with integer values, denote a configuration of our system. Starting with an initial configuration, $\vec{x}_0$, we consider a trajectory in the ACP scheme that takes it to an absorbing state, $\vec{x}_\alpha $, in $n+1$ steps, through a sequence of
configurations: 
\begin{equation}
\tau _{ACP}:\ \vec{x}_0,\ \vec{x}_{i_1},\ \vec{x}_{i_2},\ ...,\ \vec{x}%
_{i_n},\ \vec{x}_\alpha 
\end{equation}
By definition of ACP, the successive $\vec{x}_i$'s are distinct. The weight associated with this trajectory is 
\begin{equation}
p\left( \vec{x}_0,\vec{x}_{i_1}\right) p\left( \vec{x}_{i_1},\vec{x}%
_{i_2}\right) ...p\left( \vec{x}_{i_n},\vec{x}_\alpha \right) 
\end{equation}
Here 
\begin{equation}
p\left( \vec{x}_i,\vec{x}_j\right) \equiv \frac{k\left( \vec{x}_i,\vec{x}%
_j\right) }{Z\left( \vec{x}_i\right) } 
\end{equation}
where $k\left( \vec{x}_i,\vec{x}_j\right) $ is the product of the rate ($k$) and the two appropriate populations associated with the change from $\vec{x}_i$ to $\vec{x}_j$. $Z\left( \vec{x}_i\right) $ is the sum of all possible transitions, i.e., $\sum_{\vec{x}_j}k\left( \vec{x}_i,\vec{x}
_j\right) $. So, for example, $\vec{x}_i=\left( 25,144,95,319\right) $ and $ \vec{x}_j=\left( 25,145,94,319\right) $, then $k\left( \vec{x}_i,\vec{x}
_j\right) =k_b\left( 144\right) \left( 95\right) $ and $Z\left( \vec{x}_i\right) =$ $k_a\left( 25\right) \left( 144\right) +k_b\left( 144\right) \left( 95\right) +k_{c\left( 95\right) }\left( 319\right) +k_d\left(319\right) \left( 25\right) $. The overall transition probability, from $\vec{x}_0$ to $\vec{x}_\alpha $, is the sum of these weights over all such trajectories.

Turning to PRP, we can associate a class of trajectories to each $\tau_{ACP} $ 
\begin{equation}
\tau _{PRP}:\ \vec{x}_0,...\ \vec{x}_0,\ \vec{x}_{i_1},...\ \vec{x}_{i_1},\ 
\vec{x}_{i_2},\ ...,\ \vec{x}_{i_n},...\ \vec{x}_{i_n},\ \vec{x}_\alpha 
\end{equation}
where $\vec{x}_0$ is repeated $m_0$ times, $\vec{x}_{i_1}$ is repeated $m_1$ times, ... and $\vec{x}_{i_n}$ is repeated $m_n$ times. In this class, each $m$ ranges from $0$ to $\infty $. The weight associated with such a trajectory is 
\begin{equation}
\left( s_0\right) ^{m_0}\frac{k\left( \vec{x}_0,\vec{x}_{i_1}\right) }{N\left( N-1\right) /2}\left( s_{i_1}\right) ^{m_{i_1}}\frac{k\left( \vec{x} 
_{i_1},\vec{x}_{i_2}\right) }{N\left( N-1\right) /2}... 
\left( s_{i_n}\right)^{m_{i_n}}\frac{k\left( \vec{x}_{i_n},\vec{x}_\alpha \right) }{N\left(N-1\right) /2} 
\end{equation}
where 
\begin{equation}
s_i\equiv 1-\frac{Z\left( \vec{x}_i\right) }{N\left( N-1\right) /2} 
\end{equation}
is the probability for the system to stay at $\vec{x}_i$. The next step is
clear: summing over all trajectories in the class generates a factor of 
\begin{equation}
\sum_{m_i=0}^\infty \left( s_i\right) ^{m_i}=\frac 1{1-s_i}=\frac{N\left(
N-1\right) }{2Z\left( \vec{x}_i\right) } 
\end{equation}
for each $i$. This factor combines with $k\left( \vec{x}_i,\vec{x}_j\right) $
to produce $p\left( \vec{x}_i,\vec{x}_j\right) $, precisely the factor appearing in the associated $\tau _{ACP}$. In other words, each trajectory
in $\tau _{PRP}$ belongs to a class that can be associated with a unique $\tau _{ACP}$. In the limit of $t\rightarrow \infty $, the sum (over this
class) of their weights is exactly the weight for that $\tau _{ACP}$. Therefore, the transition probabilities (to go from any initial$\ \vec{x}_0$
to any particular final $\ \vec{x}_\alpha $) for PRP and ACP are identical.

\end{document}